\documentclass[prl,twocolumn,superscriptaddress]{revtex4}
\usepackage{amsmath}
\usepackage{amssymb}
\usepackage{graphicx}
\begin{document}

\title{Two energy scales in the spin excitations of the high-$T_c$ superconductor La$_{2-x}$Sr$_{x}$CuO$_{4}$}

\author{B. Vignolle}
\affiliation{H.H. Wills Physics Laboratory, University of Bristol, Tyndall Ave., Bristol, BS8 1TL, UK}
\author{S.M. Hayden}
\affiliation{H.H. Wills Physics Laboratory, University of Bristol, Tyndall Ave., Bristol, BS8 1TL, UK}
\author{D.F. McMorrow}
\affiliation{London Centre for Nanotechnology and Department of Physics and Astronomy,
University College London, London, WC1E 6BT, UK}
\affiliation{ISIS Facility, Rutherford Appleton Laboratory, Chilton, Didcot, Oxfordshire OX11 0QX, UK}
\author{H.M. R{\o}nnow}
\affiliation{Laboratory for Neutron Scattering, ETH-Z\"{u}rich and Paul Scherrer Institut, 5232
Villigen, Switzerland}
\author{B. Lake}
\affiliation{Hahn-Meitner Institut, Berlin D-14109, Germany.}
\author{C.D. Frost}
\affiliation{ISIS Facility, Rutherford Appleton Laboratory, Chilton, Didcot, Oxfordshire OX11 0QX, UK}
\author{T.G. Perring}
\affiliation{ISIS Facility, Rutherford Appleton Laboratory, Chilton, Didcot, Oxfordshire OX11 0QX, UK}

\maketitle

\textbf{
The excitations responsible for producing high-temperature superconductivity in the cuprates
have not been identified.  Two promising candidates are collective spin
excitations and phonons \cite{Chubukov2003a}.
A recent argument against spin excitations has been their inability to
explain structures seen in electronic spectroscopies such as photoemission
\cite{Bogdanov2000a,Kaminski2001a,Johnson2001a,Lanzara2001a} and
tunnelling \cite{Lee2006a}.
Here we use inelastic neutron scattering to demonstrate that collective
spin excitations in optimally doped La$_{2-x}$Sr$_{x}$CuO$_{4}$ are
more structured than previously thought.
The excitations have a two component structure with a low-frequency component strongest
around 18 meV and a broader component strongest near 40-70 meV. The second
component carries most of the spectral weight
and its energy matches structure seen in photoemission and tunnelling spectra
\cite{Bogdanov2000a,Kaminski2001a,Johnson2001a,Lanzara2001a,Lee2006a} in the range
50-90 meV. Our results demonstrate that collective spin excitations can explain
features of quasiparticle spectroscopies and are therefore likely to be the
strongest coupled excitations.}

Since their discovery, considerable progress has been made in understanding the
properties of the high-Tc cuprate superconductors.  For example, we know that the
superconductivity involves Cooper pairs, as in the conventional BCS theory,
but the $d$-wave pairing is different to the $s$-wave pairing of conventional superconductors.
However, a major outstanding issue is the pairing mechanism.  Identifying, the bosonic excitations
which were strongly coupled to the electron quasiparticles played an pivotal role in confirming
the pairing mechanism in conventional superconductors \cite{McMillan1965,Stedman1967}.
In the case of the copper
oxide superconductors, advances in electronic
spectroscopies such as angle resolved photoemission (ARPES) and tunnelling
have revealed structure in the low-energy electronic excitations which may
reflect coupling to bosonic excitations.  ARPES measurements
on Bi$_{2}$Sr$_{2}$CaCu$_{2}$O$_{8}$, Bi$_{2}$Sr$_{2}$CuO$_{6}$ and
La$_{2-x}$Sr$_{x}$CuO$_{4}$ have shown that
there are rapid changes or ``kinks''  in the quasiparticle dispersion $E(k)$
in the nodal direction [$\mathbf{k} \parallel (1/2,1/2)$] for energies
in the range 50--80 meV \cite{Bogdanov2000a,Kaminski2001a,Johnson2001a,Lanzara2001a}.
The origin of these kinks has been discussed in terms of
coupling to collective spin excitations and phonons.  Optical conductivity
measurements also suggest that there are strongly coupled electronic excitations
in this energy range \cite{Basov2005a}.  Unfortunately, the
high-energy magnetic excitations in the cuprates have been most comprehensively
studied in YBa$_{2}$Cu$_{3}$O$_{6+x}$,
a compound for which ARPES data are scarce.  In contrast,
La$_{2-x}$Sr$_{x}$CuO$_{4}$ is a system for which the magnetic excitations
may be studied by neutron scattering (because of the availability of large
single crystals) and ARPES (because a suitable surface may be prepared).

Here we report a high-resolution neutron scattering study of the magnetic
excitations in optimally doped La$_{2-x}$Sr$_{x}$CuO$_{4}$ ($x$=0.16,
$T_c$=38.5 K). The parent compound La$_2$CuO$_4$ of the
La$_{2-x}$Sr$_{x}$CuO$_{4}$ superconducting series exhibits antiferromagnetic
order with an ordering wavevector of $\mathbf{Q}_{\mathrm{2D}}$=(1/2,1/2).
Doping induces superconductivity and causes low-frequency incommensurate
fluctuations \cite{Shirane1989a,Cheong1991} to develop with wavevectors
$\mathbf{Q}_{\mathrm{2D}}$ = $(1/2,1/2 \pm \delta)$ and $(1/2 \pm
\delta,1/2)$. These excitations broaden \cite{Mason1992} and disperse inwards
initially towards (1/2,1/2) \cite{Christensen2004} with increasing energy.
The present work extends the energy range and wavevector resolution of previous studies
\cite{Shirane1989a,Cheong1991,Mason1992,Christensen2004,Hayden1996a}.

\begin{figure*}
\includegraphics[width=0.95\linewidth,clip]{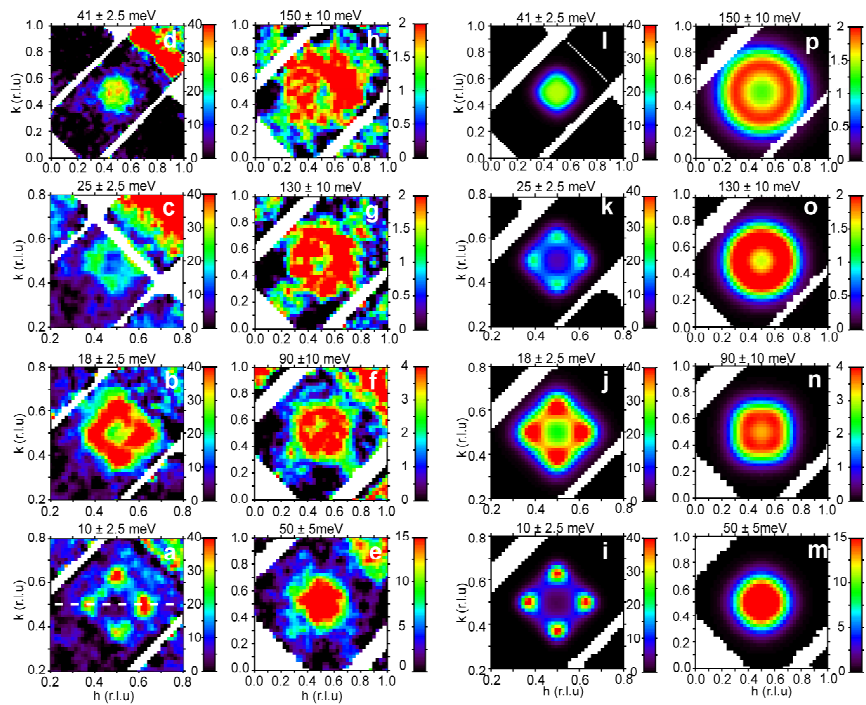}
\caption{ \textbf{Images of the magnetic excitations in La$_{1.84}$Sr$_{0.16}$CuO$_4$ for
various energies at $T$=12~K}.
\textbf{a-h} The measured $\chi^{\prime\prime}({\bf Q},\omega)$ is plotted in units of
$\mu_B^2$~eV$^{-1}$~f.u.$^{-1}$ as a function of wavevector.
\textbf{a-c} show the emergence and disappearance
of the component at
$(1/2 \pm \delta,1/2)$ and $(1/2, 1/2 \pm \delta)$ which is most intense
at lower energies.
\textbf{d-h} show the higher energy component which emerges around
41~meV and disperses outwards with increasing energy.
\textbf{i-p} Model fits to the images of the magnetic excitations
shown in \textbf{a-h}.
The phenomenological model [Eq.~(1)] provides a good description of the
experimentally measured magnetic excitations and can therefore be used to
parameterize the data.
At higher energies
(panels \textbf{n-p}) the data are best fitted with the model
$\chi^{\prime\prime}({\mathbf Q},\omega)$ rotated 45$^\circ$ in the $h-k$ plane.
Wavevectors are labelled by their positions in reciprocal space
$\mathbf{Q}=h \mathbf{a}^{*}+k \mathbf{b}^{*}+l \mathbf{c}^{*}$.
}
\label{Fig:LSCO16_MAPS_fits.eps}
\end{figure*}

Neutron spectroscopy provides a direct probe of
the magnetic response function $\chi^{\prime \prime}(\mathbf{Q},\omega)$.
Experiments were performed on the MAPS spectrometer at the ISIS spallation neutron source
of the Rutherford-Appleton Laboratory.
Figure~\ref{Fig:LSCO16_MAPS_fits.eps}a-h shows wavevector-dependent images of the magnetic response at
various energies demonstrating how it evolves with energy.
At low energies, $E=10$~meV (panel \ref{Fig:LSCO16_MAPS_fits.eps}a) we observe the low-energy incommensurate
excitations \cite{Shirane1989a,Cheong1991,Mason1992,Christensen2004}.  As the energy is increased,
$E=18$~meV (panel \ref{Fig:LSCO16_MAPS_fits.eps}b) the response becomes stronger,
the pattern fills in along the line connecting the nearest neighbour incommensurate peaks
and the incommensurability $\delta$ decreases. For $E=25$~meV (panel \ref{Fig:LSCO16_MAPS_fits.eps}c),
the intensity of the pattern is noticeably attenuated. On further
increasing to $E = 41$~meV (panel \ref{Fig:LSCO16_MAPS_fits.eps}d), the
response ``recovers'', becoming more intense again, but is now peaked at the commensurate wavevector (1/2,1/2).
At higher energy $E = 90$~meV (panel \ref{Fig:LSCO16_MAPS_fits.eps}e), the structure resembles  a
square box with the corners pointing along
the (110) type directions, {\it i.e.} towards the Brillouin zone center. Thus the square pattern
is rotated 45$^\circ$ with respect to the low energy response
(e.g. panel \ref{Fig:LSCO16_MAPS_fits.eps}b).  A similar high-energy response has been observed in
underdoped YBa$_{2}$Cu$_{3}$O$_{6+x}$
\cite{Hayden2004,Stock2005} and the stripe ordered composition La$_{1.875}$Ba$_{0.125}$CuO$_{4}$
\cite{Tranquada2004}.
Thus a rotated continuum appears to be a universal feature of the cuprates.

In order to make our analysis quantitative, we fitted a modified Lorentzian
cross-section previously used to described the cuprates and other systems \cite{Sato1974a}
to the data:
\begin{equation}
\label{Eq:Sato_Maki}
\chi^{\prime\prime}({\mathbf Q},\omega)=\chi_\delta(\omega)
\frac{\kappa^4(\omega)}
{\left[ \kappa^2(\omega)+R(\mathbf{Q})\right]^2}
\end{equation}
with
\begin{equation*}
R(\mathbf{Q})=\frac{\left[(h-\frac{1}{2})^2+(k-\frac{1}{2})^2-\delta^2 \right]^2
+ \lambda (h-\frac{1}{2})^2 (k-\frac{1}{2})^2}
{4 \delta^2}
\end{equation*}
where $\kappa(\omega)$ is an inverse correlation length, the position of the
four peaks is specified by $\delta$, and $\lambda$ controls the shape of the pattern
($\lambda$ =4 corresponds to four distinct peaks and $\lambda$=0 corresponds to a pattern
with circular symmetry).

\begin{figure}
\includegraphics[width=0.99\linewidth,clip]{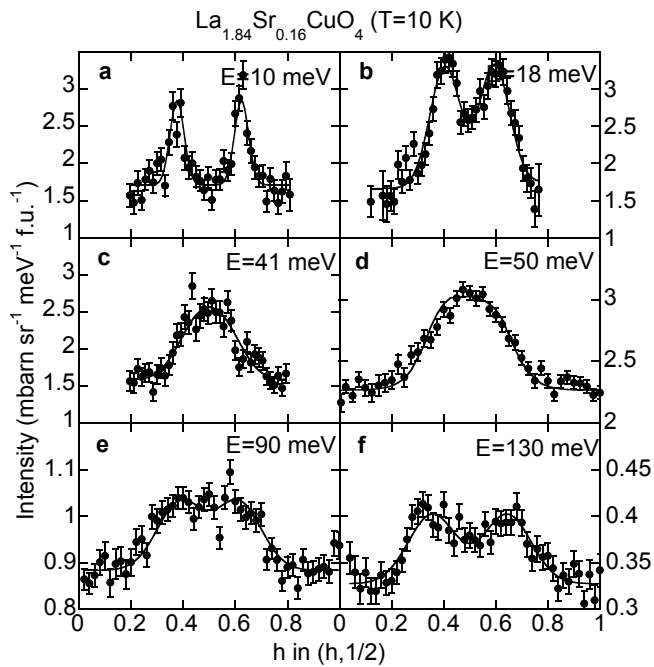}
\caption{\textbf{Magnetic excitations in La$_{1.84}$Sr$_{0.16}$CuO$_4$ at $T$=12~K.}
\textbf{a-f} show the variation of the scattered intensity with wavevector
for various fixed excitation energy.  The trajectory of the cut is
shown by the dashed line in Fig.~\ref{Fig:LSCO16_MAPS_fits.eps}a.
\textbf{a-b} show the incommensurate low-frequency component of the response.
The high-frequency component is strongest for
40--50~meV (\textbf{c-f}).  Two distinct peaks are seen at higher energies (\textbf{e-f}),
these disperse away from (1/2,1/2) in a similar manner to the spin-wave excitations in the parent compound
La$_{2}$CuO$_4$.}
\label{Fig:cuts.eps}
\end{figure}
\begin{figure}
\includegraphics[width=0.80\linewidth,clip]{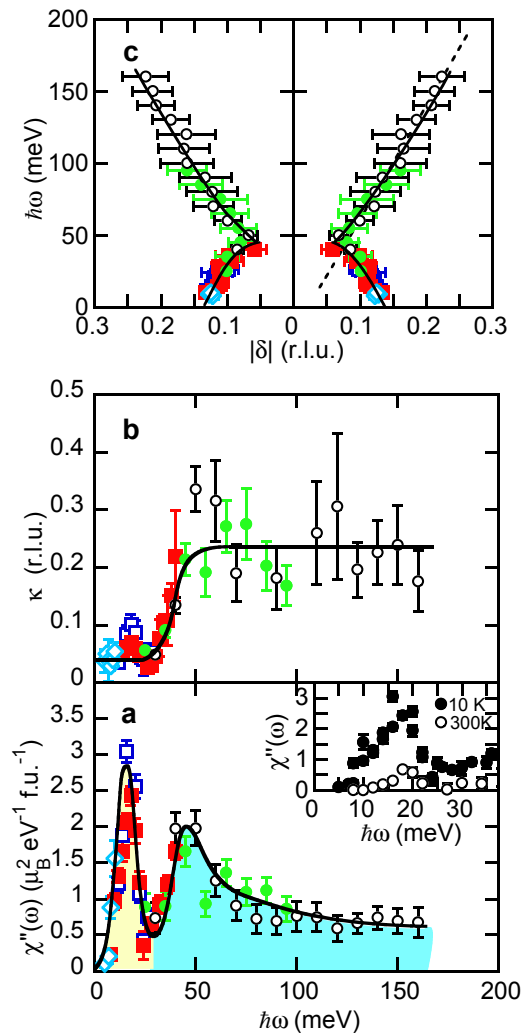}
\caption{\textbf{Magnetic excitation spectrum and evolution of the form of the magnetic response
with energy.} The susceptibility wavevector-averaged susceptibility $\chi^{\prime\prime}(\omega)$
(\textbf{a}) shows a ``peak-dip-hump'' structure suggesting that the
magnetic response has two components. The emergence of the higher frequency component
above about 40~meV corresponds to a broadening of response in wavevector as demonstrated by the
rapid increase in the $\kappa$ (\textbf{b}).  There is a strong dispersion of the peak
positions in constant excitation-energy cuts as shown by the energy dependence of the
incommensurability $\delta(\omega)$ (\textbf{c}).
The high-energy dispersion indicates the persistence of residual
antiferromagnetic interactions. Symbols indicate different
incident energies: E$_i$ = 30($\Diamond$), 55($\square$), 90($\blacksquare$),
160({\large $\bullet$}), 240 meV ({\large $\circ$}). }
\label{Fig:LSCO16_paras.eps}
\end{figure}
Fig.~\ref{Fig:LSCO16_MAPS_fits.eps}i-p shows plots of the fitted model response
for the same energies as Fig.~\ref{Fig:LSCO16_MAPS_fits.eps}a-h.
Another way of displaying the results is to take constant energy cuts through our data set.
Fig.~\ref{Fig:cuts.eps} shows cuts along the dashed trajectory in Fig.~\ref{Fig:LSCO16_MAPS_fits.eps}
for various energies together with fits of our model response (Eq.~\ref{Eq:Sato_Maki})
convolved with the experimental resolution.
The good agreement between the data and fits allows us
to use the parameters derived from the fits (Fig.~\ref{Fig:LSCO16_paras.eps}) to interpret our results.
In order to distinguish between magnetic and phonon
scattering the experiment was performed at a number of incident energies.  This
means that the same in-plane momentum $(h,k)$ could be probed with a variety of
$l$ (momentum perpendicular to plane) values and strong phonons isolated.
The fact that data collected with different incident energies yields similar results
confirms the validity of our analysis.
We have expressed the strength of the spin fluctuations in terms
of  the local or wavevector-averaged susceptibility
$\chi^{\prime\prime}(\omega)$ =
$\int \chi^{\prime\prime}(\mathbf{Q},\omega)  d^{3}Q / \int d^{3}Q$ determined
from the fitted cross-section.  The local susceptibility $\chi^{\prime\prime}(\omega)$
is a measure of the density of magnetic excitations for a given energy.

Figure~\ref{Fig:LSCO16_paras.eps}a illustrates one of the key results of this work:
the magnetic response of La$_{1.84}$Sr$_{0.16}$CuO$_4$ has a two component structure.
The lower-energy peak corresponds to the incommensurate structure which is rapidly
attenuated above 20~meV.  The
higher-energy structure is peaked at (1/2,1/2) for $E \approx$ 40--50~meV and broadens
out with increasing energy above this.  Although the wavevector-averaged susceptibility
$\chi^{\prime\prime}(\omega)$ of the higher frequency component is peaked around 50~meV,
it has a long tail with a measurable response at highest energies probed ($E$=155~meV)
in this experiment.
The observed two component structure differs from the
response observed in the magnetically ordered but weakly superconducting compound
La$_{1.875}$Ba$_{0.125}$CuO$_4$ \cite{Tranquada2004} which does not show the lower peak in
$\chi^{\prime \prime}(\omega)$.  Although La$_{1.875}$Ba$_{0.125}$CuO$_4$ does
a similar dispersion in $|\delta|$.

Given the markedly different characteristics of the two components which make up the magnetic response
in La$_{1.84}$Sr$_{0.16}$CuO$_4$,
it is likely that they have different origins.  One possible scenario is that the
lower-energy incommensurate structure is due to quasiparticle (electron-hole) pair
creation which might be calculated from an underlying band structure
\cite{Si1993,Littlewood1993}, while the higher-energy
structure is due to the residual antiferromagnetic interactions.
It is interesting to compare the magnetic response at optimal doping with that of the
antiferromagnetic parent compound La$_{2}$CuO$_4$ \cite{Coldea2001}.
In La$_{2}$CuO$_4$, $\chi^{\prime \prime}(\omega)$ is approximately
constant over the energy range probed here (0--160 meV) with
$\chi^{\prime \prime}(\omega) \approx $1.7~$\mu_B^2$~eV$^{-1}$~f.u.$^{-1}$.
Thus the effect of doping is to suppress the high-energy response
($\hbar \omega > $50~meV) and enhance the response at lower frequencies,
creating a double peak structure.
Fig.~\ref{Fig:LSCO16_paras.eps}c shows that the high energy part of the
response disperses with increasing energy. Constant energy cuts through
the data yield two peaks (see Fig.~\ref{Fig:cuts.eps}) which are reminiscent
of spin wave in the parent
compound La$_{2}$CuO$_{4}$.
We may use the high-energy dispersion to estimate an
effective Heisenberg exchange constant $J$ which quantifies the strength of the
coupling between the copper spins.  Using the fitted values of
$|\mathbf{\delta}|$ in Fig.~\ref{Fig:LSCO16_paras.eps}c for $E>\mbox{40 meV}$, we
estimate the gradient to be $dE/d\delta=510 \pm 50$~meV\AA$^{-1}$.  This may then be compared
with the standard expression for the spin wave velocity in a square lattice antiferromagnet,
$\hbar v_s = Z_c \sqrt{8} S J a$, where $Z_c$, $S$, and $a$ are the quantum renormalization,
spin, and lattice parameter respectively.  We find that the effective exchange constant
for La$_{1.84}$Sr$_{0.16}$CuO$_4$ is $J=81 \pm 9$~meV.  This is substantially reduced from the
parent compound La$_{2}$CuO$_4$ where $J=146 \pm 4$~meV \cite{Coldea2001}.

It is interesting to compare our measurements with electronic spectroscopy performed
on cuprate superconductors with the same energy scale.  The energy of the
50~meV peak matches the energy range (40--70~meV) where angle resolved photoemission
spectroscopy (ARPES) measurements
\cite{Lanzara2001a} in the same
compound La$_{2-x}$Sr$_{x}$CuO$_4$ show rapid changes or kinks in the quasiparticle
dispersion $E(k)$. These kinks may well be caused by coupling to the spin excitations
reported here.  At higher energies, ARPES measurements suggest that the quasiparticles
are coupled to bosonic excitations with energies up to at least 300~meV \cite{Kordyuk2006a}.  This would
match the tail is the collective spin excitations which we observe here (Fig.~\ref{Fig:LSCO16_paras.eps}a).
Other electronic spectroscopy data is not available  La$_{2-x}$Sr$_{x}$CuO$_4$. However, we
may compare with other systems. Infrared conductivity measurements provide evidence of coupling to high energy
excitations in, for example,  YBa$_{2}$Cu$_{3}$O$_{6+x}$ \cite{Basov1996a}
and Bi$_{2}$Sr$_{2}$Ca$_{0.92}$Y$_{0.08}$Cu$_{2}$O$_{8+\delta}$ \cite{vanderMarel2003a}.  In addition
tunnelling measurements on Bi$_{2}$Sr$_{2}$CaCu$_{2}$O$_{8+\delta}$ \cite{Zasadzinski2006a} show
evidence of coupling to a sharp mode near 40~meV and broader excitations extending above
100~meV.

In summary, we use neutron scattering measurements to show that the
collective magnetic excitations in optimally doped La$_{2-x}$Sr$_{x}$CuO$_4$ are
made up of two components.  At low energies, the excitations are incommensurate and
disperse with increasing energy towards the commensurate antiferromagnetic ordering wavevector
of La$_{2}$CuO$_4$, peaking in intensity around 20 meV. At higher energies a
second component develops which displays a broad, commensurate peak around 50 meV. It then displays
a spin-wave-like dispersion at higher energies. The high-frequency excitations are most naturally
interpreted as being due to residual antiferromagnetic interactions. Comparison of the present
data with electron photoemission and tunnelling data suggests that it is this
high-frequency component which affects the electron quasiparticles most strongly. As the
excitations that couple most effectively to the quasiparticles are most likely to play
an important part in the superconducting pairing,
our results support the notion that high-$T_c$ superconductivity is magnetically mediated.


\end{document}